\documentstyle[11pt,paspconf,epsf]{article}
\begin{document}
\title{Modeling Galaxy Lenses}
\author{Roger Blandford and Gabriela Surpi}
\affil{Theoretical Astrophysics, Caltech Pasadena, CA 91125}
\author{Tomislav Kundi\'c}
\affil{Renaissance Technologies, 600 Route 25A, East Setauket, NY 11733}
\begin{abstract}
In order to use a gravitational lens to measure the Hubble constant accurately, it is necessary
to derive a reliable model of the lens surface potential. 
If the analysis is restricted to the locations and magnifications
of point images, the derived Hubble constant depends upon the 
class of mass models used to fit the data. 
However, when there is extended emission from an Einstein ring, it may
be possible to derive a potential from the observed surface brightness
in a model-independent manner.
This procedure is illustrated with reference to B1608+656. The multi-band images are de-reddened,
de-convolved and de-contaminated so that the luminous matter and the surface brightness
contours in the Einstein ring are both faithfully mapped. This intensity distribution
can then be used to reconstruct the potential. Progress in implementing this program
is reported.

The observed incidence of multiple-imaged galaxies on the Hubble Deep Fields is an order 
of magnitude smaller than naively predicted on the basis of radio lens surveys, like CLASS,
but consistent with the rate computed using surface photometry of candidate
lens galaxies assuming standard mass to light ratios. In order to resolve this 
paradox, it is suggested that most galaxy lenses are located in compact groups.
\end{abstract}
\vspace*{-0.4cm}
\section{Introduction}
For a long while, ({\it eg} Refsdal 1964), gravitational lenses have promised unique and compelling
cosmographical measurements. Despite considerable observational progress and 
a developing theoretical sophistication, the lens community has not yet delivered on 
this promise. The largest obstacle to further progress is the modeling of the lenses. Two
novel approaches to improving our understanding of lens models are now described.
\section{B1608+656}
The well-studied quad, B1608+656 has four variable radio components arranged around an Einstein
ring and labeled A, B, C, D (Fig.~1a). The scalar magnifications relative to B of A, C, D, 
at the same emission time, are 2, 1, 0.35 and the associated delays are 26, 33, 73~d respectively.
The source and lens redshifts are known to be 0.63, 1.394 (Fassnacht, these proceedings
and references therein). The lens comprises two interacting galaxies G1, G2. Models have been 
presented in Myers {\it et al} (1995), Blandford \& Kundi\'c (1997), Koopmans \& Fassnacht (1999) 
and Fassnacht, (these proceedings). 
\begin{figure}
\vspace*{-0.8cm}
\begin{center}
\leavevmode
\epsfysize=4in
\epsfbox{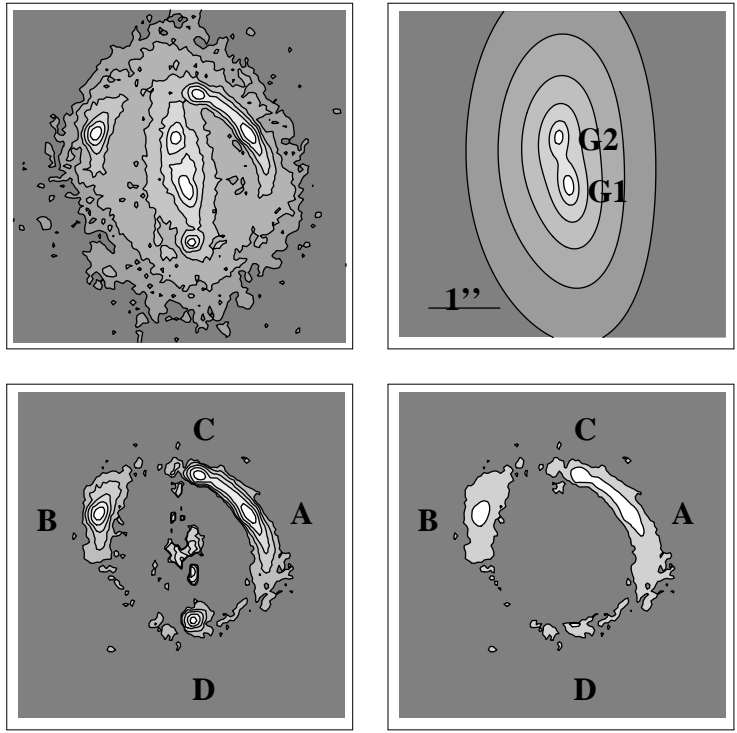}
\vspace*{-0.2cm}
\caption{B1608+656. a) V-band image from Schechter {\it et al} (in preparation) showing the four 
multiply-imaged components A, B, C, D. b) The two lensing galaxies G1, G2. The light distribution
has been corrected for reddening and the Einstein ring image of the background source has been 
removed. c) De-reddened and de-convolved I band image of the Einstein ring. The lens galaxies'
light has been removed. d) The same as c) except that ``Crossing isophotes'', 
that pass through saddle points in the intensity are highlighted.}
\vspace*{-0.5cm}
\end{center}
\end{figure}

The conventional approach to modeling galaxy lenses is to adopt a small library of potentials
or mass distributions and adopt parameters that provide the best fit to the observed
image properties by minimizing a suitably defined $\chi^2$. As more data has been acquired,
more parameters have become necessary and the accuracy of the derived value of the Hubble constant has
deteriorated ({\it eg} Barkana {\it et al} 1999). 
The fundamental problem is that when the image data is limited 
to a few isolated points there is no unique interpolation between 
them. This can be demonstrated for B1608+656 by exhibiting two different mass models that 
fit the four radio image positions and magnifications with reasonable accuracy 
as well as the ratios of the reported time delays and yet which yield Hubble constants 
of $\sim60,100$~km s$^{-1}$~Mpc$^{-1}$, respectively, (Surpi \& Blandford, these proceedings)

This degeneracy may be broken when there is extended emission 
from an Einstein ring as it is then possible to match 
points with similar surface brightness. This approach has already been 
attempted at radio wavelengths, where it
is convenient to work in Fourier space ({\it eg} Wallington, Kochanek \& Narayan 1996
and references therein). However, the method has been limited to fitting simple 
and arbitrary models of the mass distribution. 
We now discuss a somewhat different approach in which an attempt is made to solve 
directly for the surface potential from the brightness distribution and which
is specialized to address the peculiar difficulties posed by optical data.
A quite different method with a similar goal has been presented here by
Sahu (and references therein).
\begin{figure}
\vspace*{-0.8cm}
\begin{center}
\leavevmode
\epsfysize=2.5in
\epsfbox{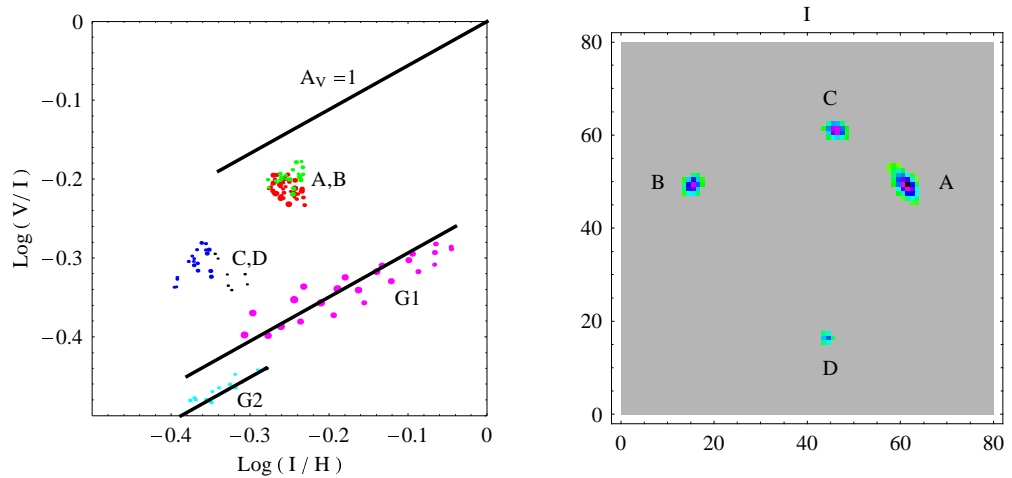}
\vspace*{-0.2cm}
\caption{Two Color Diagrams. a) The intensities of the brightest pixels in the vicinity of 
components A, B and C, D are shown, logarithmically, as ratios $V/I$ and $I/H$. The 
latter pair are displaced relative to the former by a vector parallel to the Galactic reddening
line, the upper bold line. Also shown 
are the brighest pixels in the nuclei of galaxies G1, G2 
b) The intensities of components A, B, C, D
are corrected for reddening assuming that they have similar intrinsic
colors. The effect of this transformation
is to give the four images a similar surface brightness. G1, and G2 are also corrected 
assuming a similar reddening law but separate intrinsic colors.}
\vspace*{-0.5cm}
\end{center}
\end{figure}
\subsection{Intensity Reconstruction}
We use the $V$ and $I$ band images from Schechter {\it et al}
(in preparation) and the $H$ band image from Fassnacht {\it et al}
(in preparation). These have effective wavelengths
of 372, 499, 982 nm in the lens frame, respectively. In order to form a
faithful image of the multiply-imaged source, we must 
deconvolve, de-contaminate and de-redden the observed image.
We do this by convolving 
the $V$ image with the $I$ PSF and {\it vice versa}. We then 
use these images to derive color maps of $V/I$ and $I/H$.
Next we use the observed radio magnifications and take the brightest 
$2N,N,N,0.35N$, (with $N=20$) pixels 
from images $A,B,C,D$ respectively
and plot them on a two color diagram (Fig. 2a). (The pixel numbers 
are in proportion to the radio magnifications, which we suppose 
to be unaffected by ``milli-lensing'' ({\it eg} Koopmans, these proceedings.)

We observe that  the pixels around  $A$, $B$ have similar colors and are 
presumably subject to little reddening, whereas 
those from $C$, $D$ have different colors that are displaced by a vector
similar to that associated with Galactic dust (after correcting
for the lens redshift) with $A_V=0.4,0.5$ respectively. 
(Note that we do not assume that the Galactic 
reddening law operates but can draw this conclusion from the data.) The presence of 
a constant reddening applied to the whole image will not affect our lens model;
it will affect the photometric properties of the source and lens galaxies.
\begin{figure}
\vspace*{-0.8cm}
\begin{center}
\leavevmode
\epsfysize=3in
\epsfbox{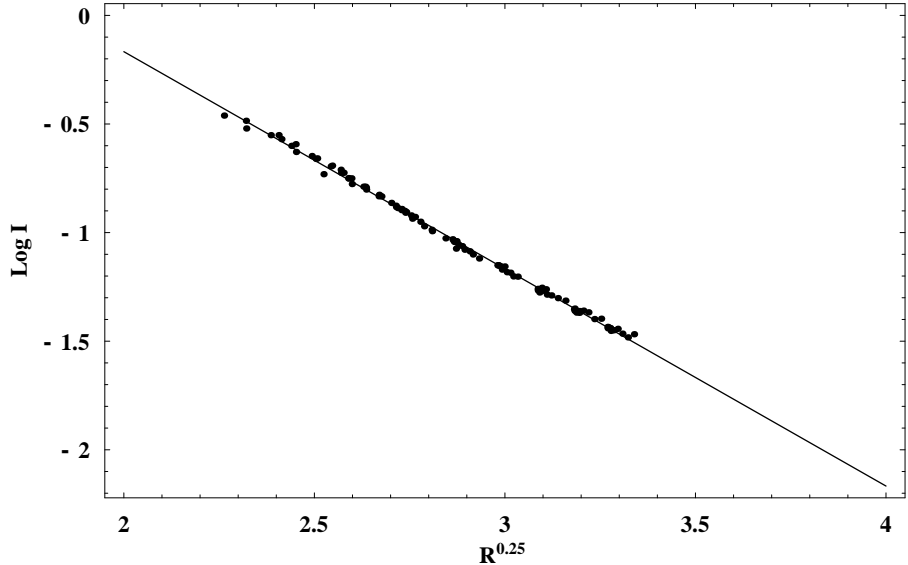}
\vspace*{-0.2cm}
\caption{De Vaucouleurs Profile fit to I band photometry of the image
of galaxy G1. Only those points which are well removed from G2 and the Einstein ring
and which are not heavily reddened are plotted.}
\vspace*{-0.5cm}
\end{center}
\end{figure}

Next we take the brightest points around the two lens galaxy nuclei, $G1$, and $G2$
and plot these on a two color diagram, we find that they lie 
along two lines also parallel to the Galactic reddening line, 
suggesting strongly that there are extinction gradients across the two lens 
galaxy nuclei.  
By inspection, we deduce that most of the 
reddening is due to $G2$, which appears to lie in front
of $G1$. If we assume that \{A, B, C, D\}, G1 and G2 have three separate
but uniform intrinsic colors, then it is possible to solve for the reddening
over most of the image. The result of de-reddening the images A, B, C, D
is shown in Fig~2b. Note that their surface brightnesses are now all similar, 
within the errors as required.

After we de-redden the lens galaxies, we observe that the surface brightness of G1,
measured at points that are well-removed from G2 and the Einstein ring,
has a  distribution with radius that matches the de Vaucouleurs profile
commonly used to describe elliptical galaxies, ({\it cf}
Kochanek, these proceedings) (Fig. 3). We assume this profile for the light 
and iteratively remove G1 to leave G2, which is too distorted 
for a simple profile to be appropriate.
In this way we can separate the light of both galaxies from that of the Einstein 
ring and, by adopting
mass to light ratios that are matched to the observed size of the Einstein
ring, we can model the luminous mass associated with the two lens galaxies. 
 
The final step is to return to the original V, I images and de-convolve the
Einstein ring using the measured PSFs. We then subtract off the 
reddened galaxy light and deredden the remaining ring image according to
the extinction map deduced above.
After iterating and some further refinements, we end up 
with a ring image like that shown in Fig.~1c. 
\subsection{Potential reconstruction}
Before we show how to use this image to solve for the surface potential we note some general
features of quad images arranged around an Einstein ring. 
We assume that the source intensity map contains a single maximum with nested,
concave isophotes. This seems to be true for B1608+656.
The challenge is to find a lens model that gives a 
four-to-one mapping of isophotes in the Einstein ring
onto isophotes of similar intensity in the source plane. 
(We ignore the fifth image near the nucleus
of G1.) Now mapping curves onto curves is not
a unique operation. (It could be made into one, if we possessed two 
sets of distinct isophotes, but we don't.) Nevertheless, there are strong constraints. 
Firstly, observe that the ``crossing isophotes'' (Fig.~1c) take
the form of three, nested ``lemniscates''
inside a ``lima\c con''. The critical curve of the lens potential reconstruction
must pass through all four saddles in the intensity, with each crossing isophote
corresponding to a simple nested isophote in the source plane that is tangent 
to the caustic. Furthermore, if we construct the ``outer limit'' 
and the ``inner limit'' curves - the loci of the two isolated image points  
associated with pairs of images merging on the critical curve - then 
these must be tangent to the crossing isophotes, as shown. These constraints
point to deficiencies in existing models.

In order to construct a surface potential, we start with a simple lens model
that distributes mass density in proportion to the derived surface
brightness in the two lensing galaxies, using a separate
mass to light ratio for each of them and extrapolating using a 
de Vaucouleurs law to large radius. It is then adjusted to locate 
the four images A, B, C, D accurately.  

This model does not yet map isophotes
onto isophotes and we must correct it. This we do by constructing 
a trial source from the average on the source plane of the image intensities.
We then map this trial source back onto the image 
plane and compare the resulting isophotes $I_1(\vec\beta)$ with the observed isophotes
$I_0(\vec\theta)$. We use the linearized equation,
\begin{equation}
I_0-I_1={\partial I_0\over\partial\vec\theta}\cdot\mu\cdot{\partial\delta\psi\over\partial\vec\theta}=
{\partial I_1\over\partial\vec\beta}\cdot{\partial\delta\psi\over\partial\vec\theta}
\end{equation}
to lowest order, where $\delta\psi(\vec\theta)$ is the correction to the normalized surface potential
and $\mu$ is the magnification tensor of the original model.
We can solve for the correction to the potential by integrating 
down a sequence of curves of steepest descent in the source plane for 
each of four image zones around A, B, C, D. The potential and its gradient
must match on the critical curve. This matching
can be accomplished, iteratively, by adjusting the intensity distribution in the source 
plane. A few iterations ought to suffice to render the model consistent with the observed 
brightness within the errors associated with the intensity reconstruction.  
In principle, we can connect A, B, C, D 
without making any assumptions about the distribution of dark matter.
\begin{figure}
\vspace*{-0.8cm}
\begin{center}
\leavevmode
\epsfysize=2.5in
\epsfbox{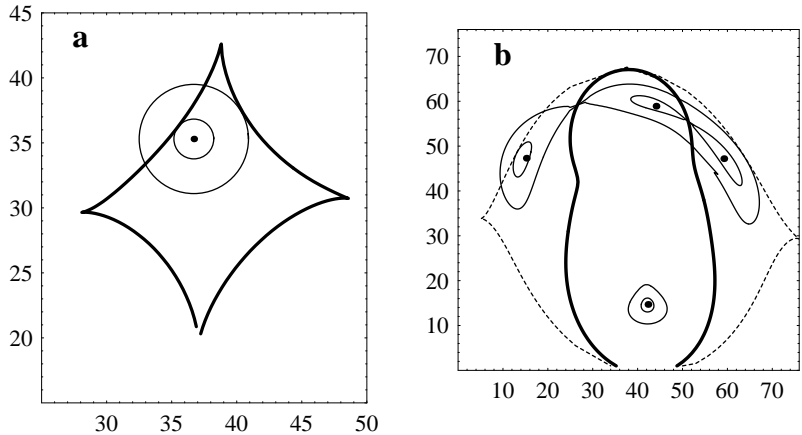}
\vspace*{-0.2cm}
\caption{Schematic Description of Lens Model a) Source plane. Source points within the astroid caustic
are associated with four image points around the Einstein ring with similar intensity
(ignoring an unobservable point near the nucleus of G1; source points
outside the caustic map onto two image points. Two circular isophotes are 
shown, chosen to be tangential to the caustic. b) Image plane. The (bold) critical curve
is the image of the caustic. The dashed curve is the outer limit. The inner limit (not shown)
is more complicated in this model but has similar properties. The isophotes, corresponding to 
the two circles on the source plane are ``crossing contours'', which must cross on the
critical curve and which are tangential to the outer (and also the inner) limit. 
The observed image imposes strong constraints on the potential.}
\vspace*{-0.5cm}
\end{center}
\end{figure}

The practical application and uniqueness of this approach 
is currently under study. 
If it is successful, we can take the Laplacian of the derived potential to give the corrected
mass distribution and subtract off the potential associated with this mass to
leave the potential associated with matter not 
covered by the Einstein ring.
Of course, this procedure conveys no information, apart from boundary condition
at the inner and outer curves, concerning 
the potential outside the Einstein ring and where the image intensity map is unreliable.

\subsection{Hubble constant}
The procedure that we have outlined can provide just the potential information that we need to convert 
the measured arrival times to a value of the Hubble constant, subject to the usual
concerns associated with the influence of intervening mass distribution and the overall
world model. The results can only be as good as the image intensity model
and the corrections made to it. It does, however, include one more internal consistency check.
The three ratios of the arrival times are determined independently by the potential model
and can be compared with the observed values. More generally, this potential reconstruction 
technique may be applicable to other extended gravitational lenses, like those associated with 
rich clusters.
\section{Strong Lensing on the HDF(N)}
\subsection{The lens deficit}
Over 15,000 radio sources have been scrutinized in the JVAS/CLASS radio surveys
(Browne, these proceedings) and they include roughly 25 confirmed gravitational lenses. 
Allowing for incompleteness etc, 
it appears that the probability of a distant radio source
being multiply imaged by an intervening gravitational lens is roughly 0.003. 
(The probability for bright quasars is somewhat larger due to magnification
bias which is less important for radio sources and faint galaxies.) Now turn to the HDF(N).
There are roughly 3000 discernible galaxy images on the roughly 5 sq. arcmin of sky
covered by the WFPC2, (giving $\sim10^{11}$ over the sky) and an expectation of 
$\sim10$ detectable cases of multiple imaging (Hogg {\it et al} 1996). There have been 
quite a few follow up observations, but there are still
no convincing examples of multiple imaging (Zepf, Moustakis \& Davis 1997, 
Blandford 1998). (One compelling case has, however, been reported
on the HDF(S), {\it cf} Barkana, Blandford \& Hogg 1999.) 

There are two immediate rationalizations of this large difference between
the radio and optical lensing rates. The first is that the faint optical galaxies are
all at very low redshift and therefore not likely to be multiply-imaged. The second is that
the HDF(N) is too small to comprise a fair sample of the lensing sky. In order to 
explore this matter further, we have carried out a more detailed analysis
of the probability of strong lensing rate ({\it cf} Blandford, 1999).
\subsection{Cross sections and multi-image probability}
Images of all of the (roughly 150) galaxies on the HDF(N) with 
spectroscopic redshifts were prepared and their rest B surface brightness converted
into surface density using mass-to-light ratios $hM/L_B=5,10$ for disk and 
elliptical galaxies respectively, together with a simple prescription for passive 
evolution ({\it eg} Vogt 1996). The galaxies were assumed to be isolated
with dark matter in their individual halos whose density declines with radius 
faster than $\sim r^{-2}$. This ensures that the surface density, which fixes
the size of the Einstein ring, is determined by the central, luminous mass.
The lack of strong color gradients 
in the galaxies of most interest suggests that reddening is not a concern.

Given these assumptions, it is possible to compute lensing cross sections
for each of these putative lens galaxies assuming that the background sources
are all at redshift $z_s=3$. The surface potentials were
computed from the surface densities using a Fourier method and 
then the total angular cross section for multiple imaging
of a point source was computed and back-projected onto the source
plane. The cross sections were all combined 
and the aggregate for the three WFPC2 chips was $\sim1$ sq. arcsec.
It was dominated by four $z\sim1$ elliptical galaxies. None of the spirals
contributed significantly to the cross section ({\it cf} Kochanek
these proceedings).  
If the HDF(N) is typical, then the multiple imaging probability per bright, distant 
source should be at most 
$\sim10^{-4}$, over an order of magnitude smaller than suggested by 
the radio surveys. 

One worry about this result is that the multiple images might actually be hidden
in the cores of the lensing galaxies and have not been recognized as such.
This was checked observationally by imaging actual
faint galaxies taken from the HDF(N)
through the principal lens galaxies. It turned out that the magnified 
source population was generally recognizable. (It is much easier to see a multiply imaged
faint galaxy through an elliptical lens than through a spiral.)
Actually as a result of this investigation, it was found that one of the elliptical 
galaxies contributing significantly to the total cross section did contain a faint,
arc-like feature in its nucleus, possibly a merger, but conceivably a lens. 
Either way it does not change the conclusion that the 
total cross section for strong lensing 
over the area of sky covered by the HDF(N) is ten times 
smaller than average. 
\subsection{Lensing by Groups}
When {\it bona fide} radio lenses are examined in detail, it is found that 
several of them have companion galaxies that are almost certainly
contributing to the imaging. Furthermore, upon spectroscopic examination,
several of these companion galaxies have similar redshifts
to the nominal lens galaxy ({\it eg} Kundi\'c {\it et al} 1997ab, Lubin 
{\it et al} 2000 in press). This suggests that a good fraction  of the radio lens 
galaxies are ellipticals belonging to compact groups. This inference
is consistent with the conclusion of a pencil beam redshift survey of 
$z\sim0.5-1$ field galaxies which shows that most of the ``absorption''
line galaxies are in compact redshift groupings Cohen {\it et al} (1999).

Groups probably form within substantial dark matter perturbations.
Although the surface density of the dark matter alone may not exceed the 
critical value, it may well be sufficient to enhance the cross section
and the size of the Einstein ring in those elliptical galaxies that 
are located near the centers of the richest and most compact groups, 
({\it eg} Zabludoff \& Mulchaey 1998, Mulchaey \& Zabludoff 1998).

Let us make a simple model of a giant elliptical galaxy located at
the center of a compact group. We suppose that the dark matter in the group
is centered on the galaxy and has a profile, 
$\rho=\rho_{\rm gp0}(1+r^2/s_{\rm gp}^2)^{-3/2}$. The galaxy is taken to have 
a density profile $\rho=\rho_{\rm gal0}(1+r^2/s_{\rm gal}^2)^{-1}$, that is to say it
is isothermal in its outer parts which extend to a tidal radius $r_{\rm tid}
\sim0.5s_{\rm gp}\sigma_{\rm gal}/\sigma_{\rm gp}$ where its density 
matches that of the group. (We assume that the group velocity dispersion
$\sigma_{\rm gp}$ is larger than that in the outer parts of the galaxy 
$\sigma_{\rm gal}$.)

The cross section to multiple imaging can be approximated by the solid angle
subtended by the Einstein and a straightforward calculation
furnishes the estimate
\begin{equation}
\pi\theta_E^2={\pi s_{\rm gal}^2\over D_d^2}\beta(\beta-2)
\end{equation}
where 
\begin{equation}
\beta={4\pi\sigma_{\rm gal}^2D_dD_{ds}\over(1-A)s_{\rm gal}D_s}
\end{equation}
(with $A=0$), is a measure of the lensing strength of the galaxy and 
\begin{equation}
A={18\sigma_{\rm gp}^2D_dD_{ds}\over s_{\rm gp}D_s}
\end{equation}
measures the extra magnification associated with the dark matter in the group.
Numerically, and very roughly, for $z_d\sim0.5,z_s\sim2$, 
$\sigma_{\rm gal}\sim200$~km s$^{-1}$, $s_{\rm gal}\sim2h_{60}^{-1}$~kpc, say, then
the cross section for an isolated elliptical galaxy like one of those observed
on the HDF(N) is $\sim0.5$~sq arcsec and $\beta=3$. Now, if 
$\sigma_{\rm gp}\sim $~500km s$^{-1}$ and $s_{\rm gp}\sim100h_{60}^{-1}$~kpc, then 
$A\sim0.5$, $\beta$ doubles and the cross section increase by a factor
24 to $\sim7$~ sq arc sec, comparable with that observed in group 
lenses. Although this example is very simple-minded, it does illustrate a general
point, namely that the cross section of an elliptical galaxy can be very
sensitive to the presence of dark matter in a surrounding group.  

Returning to the HDF(N), it appears that it contains (perhaps through selection)
no massive elliptical - group combinations with the most propitious 
redshifts for lensing, $z\sim0.5$.  The probability that a
particular elliptical - group be aligned with a suitable source 
and produce a prominent optical ring is, in any case, typically less than $\sim0.3$,
even at intermediate redshift. It is therefore not unreasonable that
no lenses have been seen. A larger area of the sky must be imaged to
the depth of the HDF(N) to have a fair sample. We do not yet know the redshift
distribution of the faint source galaxy population (though this can be 
ascertained by weak galaxy-galaxy lensing and strong cluster lensing)
but it is not required that most faint galaxies are local.

There are three consequences of this interpretation. Firstly, as already 
reported by Keeton, Kochanek \& Falco (1997), lens galaxies should 
exhibit larger than average mass-to-light ratios. Secondly, the dark
matter groups that we postulate to enhance the cross section should be 
detectable locally as X-ray sources. The matter density in this form
has a cosmological density which we estimate to be 
$\sim0.03$, roughly ten percent of the total. Thirdly, accurate
modeling of the lenses, as we have attempted for B1608+656,
should actually require the addition of asymmetric dark matter and this
may account for the unusually high proportion of quads. Of course not
all galaxy lenses are located in groups or are associated with elliptical 
galaxies, but it is our contention that a significant fraction 
will turn out to be so.

A fuller treatment of these ideas will be presented elsewhere.
\acknowledgements
This research owes much to the careful radio, infrared and optical observations of B1608+656
led by Chris Fassnacht, Tony Readhead and Paul Schechter, respectively. 
Tereasa Brainerd, Judith Cohen, David Hogg and Lori Lubin 
are thanked for collaboration on parts of the HDF analysis. 
Support under NSF grant AST is gratefully acknowledged. 
RB thanks the Institutes for Advanced Study, 
of Astronomy and of Theoretical Physics for hospitality and
NSF (through grant AST99-00866) for support, respectively.
\vfill\eject

\end{document}